\documentclass{ieeeaccess}
\usepackage{cite}
\usepackage{amsmath,amssymb,amsfonts}
\usepackage{algorithmic}
\usepackage{graphicx}
\usepackage{textcomp}
\def\BibTeX{{\rm B\kern-.05em{\sc i\kern-.025em b}\kern-.08em
    T\kern-.1667em\lower.7ex\hbox{E}\kern-.125emX}}
\begin{document}
\history{}
\doi{10.1109/ACCESS.2023.3314700}
\title{Deep learning based projection domain metal segmentation for metal artifact reduction in cone beam
computed tomography}
\author{\uppercase{Harshit Agrawal}\authorrefmark{1,2},
\uppercase{ Ari Hietanen\authorrefmark{2}, and Simo S\"{a}rkk\"{a}}\authorrefmark{1},
\IEEEmembership{Senior Member, IEEE}}
\address[1]{Department of Electrical Engineering and Automation, Aalto University, 00076 Aalto, Finland}
\address[2]{Department of Research and Technology, Planmeca Oy, 00880 Helsinki, Finland}
\tfootnote{The work was supported in part by a Business Finland grant.}
\IEEEoverridecommandlockouts
\IEEEpubid{\makebox[\columnwidth]{978-1-4799-7492-4/15/\$31.00~
\copyright2015
IEEE \hfill} \hspace{\columnsep}\makebox[\columnwidth]{ }} 
\markboth
{Author \headeretal: Preparation of Papers for IEEE TRANSACTIONS and JOURNALS}
{Author \headeretal: Preparation of Papers for IEEE TRANSACTIONS and JOURNALS}

\corresp{Corresponding author: Harshit Agrawal (e-mail: harshit.agrawal@aalto.fi)}

\begin{abstract}
Metal artifact correction is a challenging problem in cone beam computed tomography (CBCT) scanning. Metal implants inserted into the anatomy cause severe artifacts in reconstructed images. Widely used inpainting-based metal artifact reduction (MAR) methods require segmentation of metal traces in the projections as a first step, which is a challenging task. One approach is to use a deep learning method to segment metals in the projections. However, the success of deep learning methods is limited by the availability of realistic training data. It is laborious and time consuming to get reliable ground truth annotations due to unclear implant boundaries and large numbers of projections. We propose to use X-ray simulations to generate synthetic metal segmentation training dataset from clinical CBCT scans. We compare the effect of simulations with different numbers of photons and also compare several training strategies to augment the available data. We compare our model's performance on real clinical scans with conventional region growing threshold-based MAR, moving metal artifact reduction method, and a recent deep learning method. We show that simulations with relatively small number of photons are suitable for the metal segmentation task and that training the deep learning model with full size and cropped projections together improves the robustness of the model. We show substantial improvement in the image quality affected by severe motion, voxel size under-sampling, and out-of-FOV metals. Our method can be easily integrated into the existing projection-based MAR pipeline to get improved image quality. This method can provide a novel paradigm to accurately segment metals in CBCT projections.
\end{abstract}


\titlepgskip=-15pt

\maketitle

\section{Introduction}
\label{sec:introduction}
\PARstart{C}{one} beam computed tomography (CBCT) acquires 2D X-ray projections of an object to reconstruct 3D images. The presence of high-density metallic implant in the object may corrupt the projection data and cause severe artifacts such as blooming, streaking, and shading \cite{schu:11} in the reconstructed images. Several projection-based metal artifact removal (MAR) methods have been proposed in literature \cite{Kal:87, Mey:10, Wang13R, MEILINGER11, xin08, peng17}. A crucial step in these methods is the segmentation of corrupt metal traces in the projections. First, the metals are segmented in the 3D image-domain by a threshold. Then the segmented metals are forward-projected to obtain the metal trace in the projections. The data inside the metal trace is discarded and interpolated from the nearby projection data. The corrected projections are back-projected (reconstruction) to generate corrected 3D images.

Several deep learning-based MAR methods have been proposed recently. They either work in projection-domain, image-domain, or both. Projection-domain methods \cite{Hao:19, Ghani20, peng20, Agrawal21, Jun21} inpaint the metal trace in the projection (CBCT) or sinogram (CT) using deep learning architectures. All of these methods rely on the availability of segmented metal traces. Image-domain methods, such as \cite{Gjesteby17, Liang19} directly translate from artifact to artifact-free images. Image-domain methods can effectively reduce metal artifacts in some cases, however, their effect can be limited by the presence of scatter, large and varying shapes of metals, multiple metal implants, and motion artifacts \cite{Ghani20}. While inpainting-based methods deal better with these problems, they still require accurate metal segmentation in the projection \cite{Moseley05}. In \cite{peng22, yu21, lin19}, it was proposed to incorporate both image and sinogram-domain deep learning. These methods use filtered back projection (FBP) and forward projection (FP) layers to propagate the loss from the image-domain to  the sinogram-domain. These methods were implemented for CT images (slice-wise reconstruction) though full volume CBCT reconstruction is not feasible due to its large memory requirements \cite{ketcha21}.

It was demonstrated in \cite{St13} that precise segmentation of metal traces can enhance the image quality, and conversely, less accurate segmentation may cause additional artifacts or even remove anatomical details. As forward projection of the image-domain metal is used to obtain the metal traces in the projections, it has several problems. When metal lies outside of the field-of-view (FOV), image-domain segmentation methods cannot segment metals in the projections \cite{Seu:20}. This may cause image inhomogeneity in the center and among the different zones of the FOV \cite{Ama:18,Ama:20}. Furthermore, in the presence of motion artifacts \cite{nardi16, Agrawal2018, man99}, the metals are blurred and it is difficult to obtain reliable forward projections of the image-domain segmented metals. Pre- and post-processing steps were proposed in \cite{tof14, hahn18} to mitigate motion affected metal artifacts. Moreover, the forward projection step of the metals need to account for the partially covered pixels. This can be done by reconstructing the voxels with smaller size before forward projection or by taking the voxel diameter into account \cite{Amirkhanov11} at the cost of increased complexity and processing time.
 
While several deep learning methods have been investigated for MAR, few have been proposed to directly segment the projection-domain metals in CBCT scans, owing to the unavailability of accurate metal labels \cite{Hega19, Got21}. A U-Net \cite{Ronneberger15} architecture was used in \cite{Hega19} to segment metals in dental CT projections. The network was trained and tested on a very small dental CT dataset (five patients for training and four patients for testing) due to the time consuming process of creating ground truth labels. Pairs of metal and metal free projections from cadaver CBCT scans were used in \cite{Got21} to train a U-Net architecture. Acquiring data from cadaver images is a rare process and the data might still lack the quantity and variety of images. They used a consistency check condition to reduce false positives which provides more consistent segmentation of metals in the projections. The consistency check involved extra steps to reconstruct larger volumes and needed to calculate accurate thresholds for the metals in the reconstruction. Furthermore, we found that consistency check did not work well for motion-affected scans.

Monte Carlo simulations have been used to generate training data for deep learning-based metal in-painting and scatter correction methods \cite{yu21, peng20, Maier19}, but the use of simulated dataset in metal segmentation training has not been investigated yet. Motivated by the importance of accurate training data for metal segmentation, we used Monte Carlo simulations to generate metal-corrupted CBCT projections and corresponding metal labels for network training. For the tasks of in-painting and scatter correction, simulations with low noise levels are needed \cite{yu21, peng20, Maier19}. The noise in the simulated projections decreases with the increase in numbers of photons per detector pixel \cite{Maier19}, however, running the simulations with large number of photons is computationally expensive. In contrast, for metal segmentation training task, we used simulations with only 300 to 1400 photons per detector pixel (noisy) and compared the results with the simulations of 5000 photons per detector pixel (clean). Noisy simulations are faster to compute and add realistic noise to the dataset.

We trained a modified version of U-Net only on the simulated data to segment metal traces in the projections. In addition, we used a simple strategy to augment our simulated dataset further by training the network on full size and cropped projections together. Our simulated dataset included multiple anatomies such as knee, wrists, ankle, palm, and foot. We demonstrate the robustness of our metal segmentation model on 10 metal-affected and 6 non-metal clinical scans. 

As U-Net architectures have been used in \cite{Hega19, Got21} for the metal segmentation in projection data, using U-Net is not a contribution of this work. However, our main contributions are as following:

\begin{enumerate}
\item We propose a new training approach for metal segmentation in CBCT projection that uses simulated dataset (with different noise levels) obtained from real metal-affected cases. 

\item We propose a simple data augmentation strategy by training the network with crops and full size images. This strategy effectively reduces false positives in 6 unseen clinical CBCT scans without metals. 

\item We demonstrate the metal artifact reduction on 10 unseen clinical CBCT scans affected by metals. The results show good performance in challenging cases such as motion-affected and out-of-FOV metals, and they are robust to voxel size changes. 
\end{enumerate} 

\section{Methods}
In this section, we explain the dataset used for the experiments, our pipeline to generate the training data, the simulation process, the network architecture used for segmentation of metals, implementation of the network, and the evaluation methods in detail.

\begin{figure*}[tbh]
{\includegraphics[width=1\textwidth]{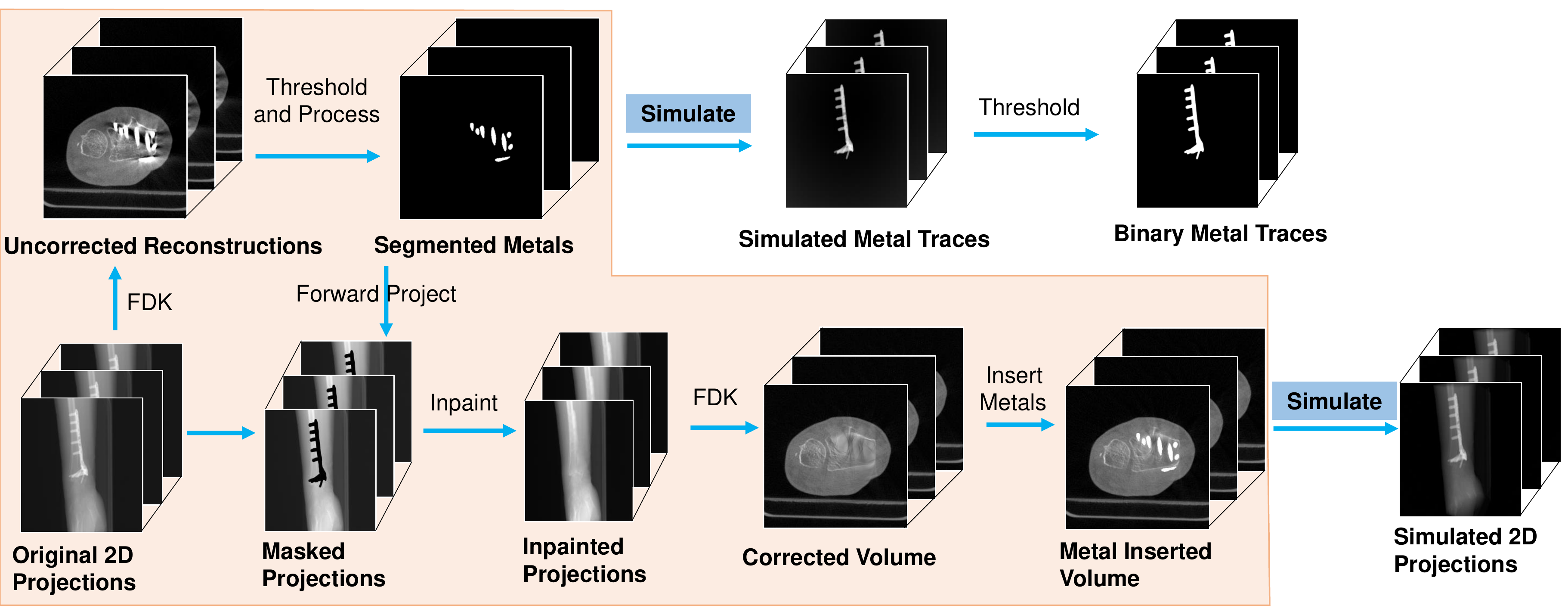}}
\caption{Pipeline for generating training pairs with examples of intermediate results. The yellow colored box illustrates conventional method for metal artifact reduction. The metal corrupted original 2D projections were first reconstructed using FDK \cite{Feldkamp:84}. The metal was segmented and refined. The metals are put back in the initial corrected volume. Then metal and metal inserted volumes were simulated using Monte Carlo simulations. The ground truth binary traces were obtained by applying a threshold to the simulated metal masks.}
\label{pipeline}
\end{figure*}
\subsection{Dataset}

We used an unidentifiable and pseudonymized clinical dataset acquired by Planmed Verity{\textregistered} scanner (Planmed Oy, Helsinki, Finland) with patient permission for use in research and development studies. This study was performed in line with the principles of the Declaration of Helsinki. The dataset had 26 clinical scans taken at different extremity anatomies and was collected over multiple years, which makes the dataset diverse and realistic. Out of 26 scans, 20 were metal-affected and 6 were without any metal. Out of 20 metal-affected scans, we included 10 scans for simulated training data generation. The remaining 10 metal-affected and 6 non-metal scans were used for evaluation of the trained models. Source-to-detector distance was 580 mm and source-to- isocenter distance was 392 mm. The description of the training and testing dataset is given in Table \ref{data}. Each scan had multiple projection views, ranging from 300 to 450. The tube voltage varied from 90 kV to 96 kV. All scans had isotropic spatial resolution, with voxel spacings of 0.6mm, 0.4 mm or 0.2 mm. Different anatomy locations were scanned, including, knee, wrist, foot, ankle, palm, and forearm. The detector size was $238.76$ mm $\times$ $189.99$ mm. The size of each detector pixel was $0.254$ mm $\times$ $0.254$ mm. The dimension of each projection was $948 \times 740$ pixels. 

\begin{table}[hbt!]
\label{tabel 1}
\caption{Description of scans used for simulated training data generation and testing of the trained models. kV is the tube voltage in kilovolt, voxel is the size of reconstructed voxels in mm, views are the number of projections per scan and anatomy is location of the body part irradiated with X-rays. kV values were not available for set 1 and set 2 in train set.}
\centering
\setlength{\tabcolsep}{0pt} 
\begin{tabular*}{\columnwidth}{@{\extracolsep{\fill}\quad}ccccccccc}
\hline
\multicolumn{1}{c}{}
& \multicolumn{4}{c}{Train}
& \multicolumn{4}{c}{Test} \\
\hline
{} & {kV} & voxel & views & anatomy & {kV} & voxel & views & anatomy \\
\hline
{set1} & 96 & 0.4 & 400 & knee & 92 & 0.2 & 300 & foot\\
\hline
{set2} & - & 0.2 & 300 & wrist & 96 & 0.2 & 400 & ankle\\
\hline
set3 & - & 0.2 &  300& ankle & 96 & 0.4 & 400 & ankle\\
\hline
set4 & 96 & 0.2 & 300 & knee & 96 & 0.2 & 300 & leg\\
 \hline
set5 & 96 & 0.2 & 450 & foot & 90 & 0.2 & 300 & forearm \\
\hline
set6 & 90 & 0.2 & 300 & wrist & 90 & 0.4 & 400 & knee\\
\hline
set7 & 90 & 0.2 & 300 & wrist & 96 & 0.2 & 400 & foot \\
\hline
set8 & 90 & 0.2 & 300 & wrist & 96 & 0.6 & 400 & leg \\
\hline
set9 & 96 & 0.2 & 300 & ankle & 92 & 0.2 & 300 & knee \\
\hline
set10 & 90 & 0.2 & 300 & palm & 90 & 0.2 & 450 & wrist \\
\hline
\end{tabular*}

\label{data}
\end{table}

From the training set of 10 scans in Table \ref{data}, we created 10 reconstructions. Two more reconstruction were created by combining metals from multiple reconstructions. These reconstructions were used to generate 3450 projections using simulation procedure described in Section \ref{ssec:simulation}. As we did not want to compromise the quality of segmentation near metal boundaries, we did not downsample projections for training as in \cite{Hega19}. To increase the data variability, we created four crops of size $474 \times 370$ from each projection. The crop was included in the training only if the sum of metal pixels was at least 100 pixels in the cropped projection. Thus we had 9305 crops for training. 
From the test set of 10 metal-affected scans in Table \ref{data}, we created metal labels manually. Since it was time consuming to segment metals in all of the projections, we manually segmented metals in 10 projections from each of the 10 scans. So, in total we gathered 100 pairs of metal-corrupted projections and corresponding ground truth metal labels from the clinical dataset.
For the remaining test set of 6 metal-free scans, we did not require ground truth labels. Those scans had a total of 2400 metal-free projections.

\subsection{Training-Data Generation Pipeline}
\label{ssec:training-data}

The pipeline to generate pairs of metal corrupted  projections and corresponding ground truth metal traces is shown in Fig.~\ref{pipeline}. The pipeline had two main parts. In the first part (shown in the colored box), a segmentation of metals in 3D and an initially corrected volume were obtained. In the second part, a simulation process was used to generate pairs of training data which is described in detail in Section \ref{ssec:simulation}.

First, the 3D image volume was reconstructed from real CBCT projections. Then we segmented the image-domain metals in the 3D reconstructed images using a global threshold. The segmented metals were not smooth and complete due to the artifacts remaining in the reconstruction. To smoothen and complete the metal boundaries, we applied median filtering and binary dilation to the segmented metals in 3D. The segmented 3D metals were further clipped to a minimum value of 3400 Hounsfield Units (HU). The segmented 3D metals were simulated to get only-metal projections. Binary metal traces for the ground truth were obtained by applying a threshold to the simulated metal projections. This thresholding step was straightforward and was done to assign a value of 1 to the metal pixels and 0 to the background. The binarized metal traces constituted ground truths for the model training.

\begin{figure}[tbh]
\centerline{\includegraphics[width=\columnwidth]{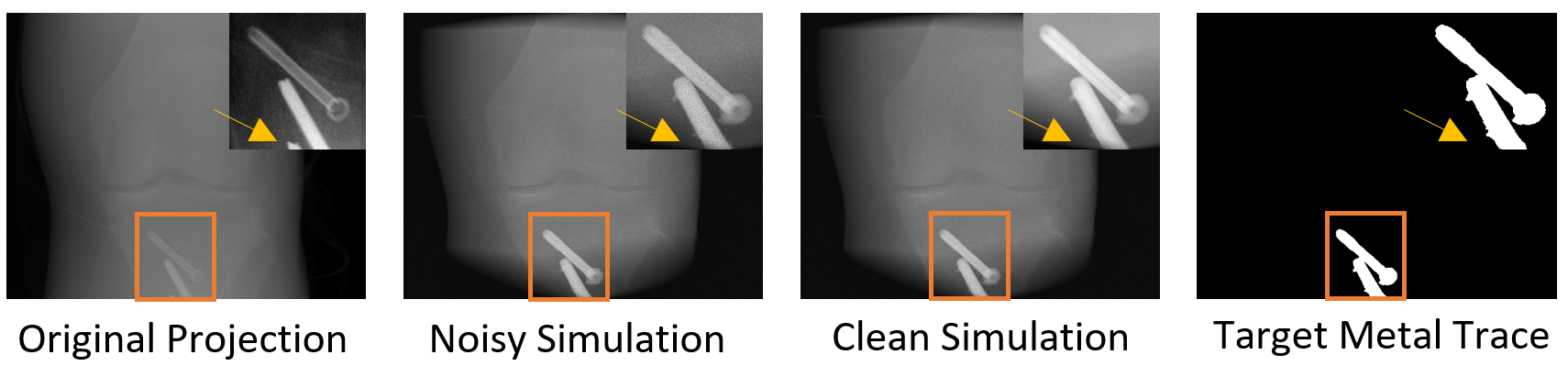}}
\caption{An example of original projection, simulated projections, and the corresponding target metal trace. The noisy projection was generated with 300 photons per detector pixel and the clean projection was generated with 5000 photons per detector pixel. The area around metals is enlarged to show the differences. The simulated projections and the corresponding mask were used for training.}
\label{sim}
\end{figure}

To obtain the metal corrupted projections, we first created metal free reconstruction. The 3D metals obtained from the first step were put back into the metal free reconstruction and the metal-inserted volume was used to simulate metal-corrupted projections. An example of original projection, a simulated projection and the corresponding ground truth metal trace is shown in Fig.~\ref{pipeline}. The simulated ground truth metals do not always align with the original projection due to morphological operations and inaccuracies in the 3D metal segmentation step. However, the metal ground truth aligns accurately with the metals in the simulated metal corrupted projection. An example of such simulations is shown in  Fig.~\ref{sim} where third small metal is not present in the traget metal mask and the two metals are dilated in comparison to the metals in the original projection. However, the metals in the target metal trace aligns well with the simulated projections. This is the reason why we used simulated projections for the training instead of original projections.

\subsection{Simulation Process}
\label{ssec:simulation}
 We used Monte Carlo simulations to generate accurate metal segmentation training dataset. The implementation was similar to the one described in \cite{RN50, badal2009accelerating}. We simulated path of X-ray photons through  the voxelized reconstructions obtained from 10 metal-affected clinical CBCT scans as described in Section \ref{ssec:training-data}. Such clinical scans represent realistic variations in the metal shapes and sizes. The reconstruction volumes were first segmented into four different materials by applying thresholds on HU values. The volumes were segmented into air, soft tissue, bone, and metal as follows:
\begin{equation}
\mathbf M_{x,y,z} = \begin{cases}
			\text{air}, & \text{if} \: \mathbf V_{x,y,z} < -500,\\
            \text{soft tissue}, & \text{if} \: -500 \leq \mathbf V_{x,y,z} < 500, \\
            \text{bone}, & \text{if} \: 500 \leq \mathbf V_{x,y,z} < 3400, \\
            \text{metal}, & \text{if} \: \mathbf V_{x,y,z} \geq 3400,
		 \end{cases}
\label{eq1}
\end{equation}
where $\mathbf V_{x,y,z}$ is the voxel value in Hounsfield Units (HU) at a location $x,y,z$ in the reconstructed volume $\mathbf{V}$ and  $\mathbf M_{x,y,z}$ is the corresponding material obtained from thresholds.
 
\begin{figure}[tbh]
\centerline{\includegraphics[width=\columnwidth]{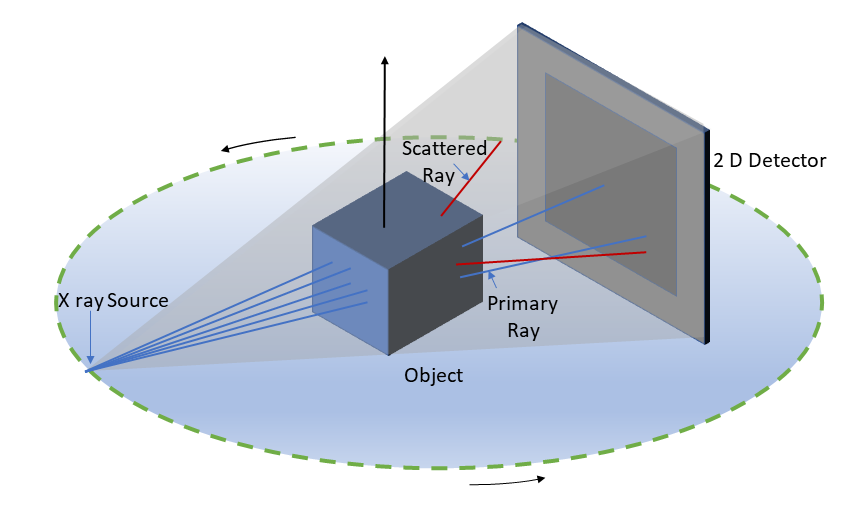}}
\caption{Illustration of CBCT geometry for Monte Carlo simulations. Incident rays leaving the object along a straight line  primary (blue) and scattered rays (red) are shown.}
\label{simD}
\end{figure}

 We simulated the voxelized geometries of CBCT scanner used to acquire the clinical scans to keep the dataset realistic. Each X-ray photon was started from a point source located at the position given by the scanner geometry. The path of the photon was simulated along a straight line from the X-ray source location to the rectangular detector grid as shown in Fig. \ref{simD}. The initial energy of each photon was randomly sampled from a polychromatic energy distribution according to the kV used during the scan. When the X-ray photon passes through the object volume, the photon interacts with the object's material according to object's density and photon energy. Three types of X-ray interactions were simulated, namely, photoelectric absorption, Compton scattering, and Rayleigh scattering based on \cite{Cullen97}. Energy-dependent attenuation curves provided in \cite{Cullen97} were used for each of the material obtained from the segmentation in \eqref{eq1} to model interactions of the polychromatic X-ray photons during transport. Projections were formed by integrating the energy entering detector pixels. More detailed explanation about the Monte Carlo simulations can be found in \cite{RN50, badal2009accelerating}.
We simulated multiple 2D projection views for each static reconstruction volume by rotating the source and detector according to the original geometry used by CBCT scanner. Furthermore, a simulation was done for each projection view without the object in the field-of-view to obtain 2D projections, called flat field. Projections were normalized and linearized, according to
 \begin{equation}
    p_\text{sim} =  -\log \left(\frac{S}{F}\right),
\end{equation}
where $S$ is the simulated projection and $F$ is the simulated flat field projection. 

\subsection{Network Architecture}
\label{sec:unet}
\begin{figure*}[tbh]
{\includegraphics[width=\textwidth]{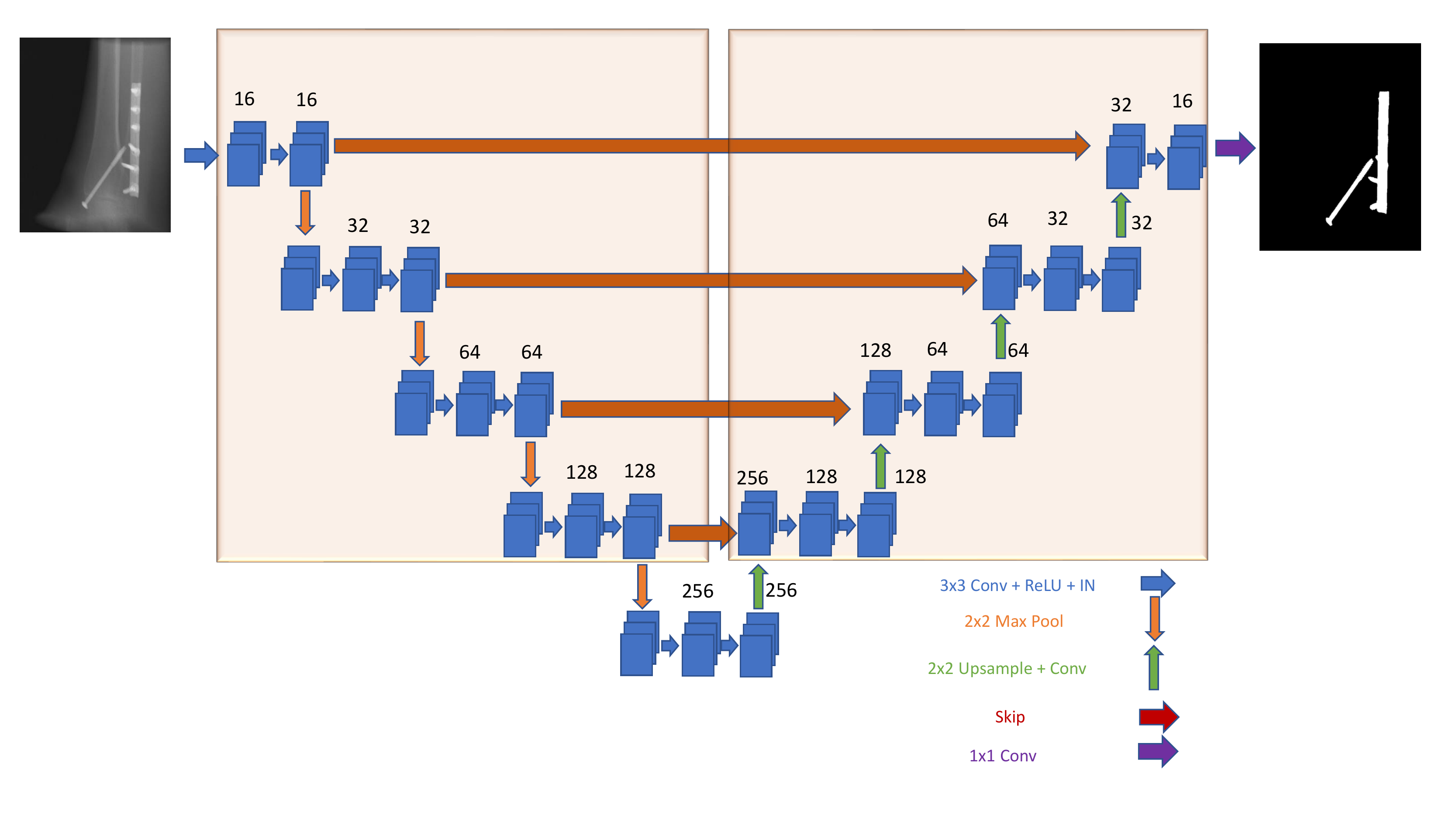}}
\caption{Network architecture used for metal segmentation. Simulated projections are  given as input to the network. A sigmoid function and a threshold of 0.5 was applied at the output of the network to get binary metal trace.}
\label{unet}
\end{figure*}

The modified U-Net architecture used for metal segmentation training is shown in Fig.~\ref{unet}. It consisted of 4 downsampling (encoder) and 4 upsampling (decoder) blocks. Each downsampling block had two convolutional layers (kernel size = $3 \times 3$, stride = $1 \times  1$, zero padding), each followed by a rectified linear unit (ReLU) and  an instance normalization layer \cite{UlyanovVL16}. In the first block, the network had 16 channels. The number of channels was doubled in each block. After every two convolutional layers, a $2 \times 2$ max-pooling layer was applied to downsample features. Each upsampling block had a bilinear upsampling layer followed by a convolutional layer (kernel size = $3 \times 3$, stride = $1 \times 1$, zero padding). The features from encoder blocks were concatenated to the features of decoder blocks to allow flow of high resolution information from earlier layers. Output of each upsampling block was padded to match the dimensions of the skip connection from the encoder side. This padding scheme ensured the size of input and output remain same and varying sizes of input images could be given to the network during training. The final layer in the network was a $1 \times 1$ convolutional layer without any activation. A sigmoid function was applied after the final layer of the network to convert the network output to a probability map. To obtain metal segmentation during the inference, pixels with probability equal to or above 0.5 were classified as metal and pixels below the probability of 0.5 were classified as background. 

\subsection{Implemetation Details}
Our network was implemented using PyTorch \cite{py15} deep learning library and was trained on a single GeForce RTX 2080 Ti Rev. A (11GB), with a batch size of 4. We used Adam optimizer ($\beta_{1}=0.9, \beta_{2}=0.999$) \cite{Adam} to optimize the network parameters to minimize loss between the network output and ground truth. The initial learning rate was set to  $e^{-4}$ and reduced after each epoch in logarithmic steps down to $e^{-6}$. The learning rate was reduced during the first 25 epochs and was fixed to $e^{-6}$ after that. As \cite{zhang} showed that stacked data augmentations are effective for the model's generalization in medical segmentation tasks, we applied a number of stacked data augmentations during training. Each input projection was normalized by its maximum value before augmentation. We used horizontal and vertical flips, image shift, rescale, rotation, Gaussian noise, multiplicative noise, elastic transformation, and mask dropout augmentations. The augmentations were applied on-the-fly during training with a probability of 0.2 for each augmentation. Therefore, each epoch had a slightly different set of inputs. To avoid over-fitting, the training was stopped if the loss did not decrease for five epochs (early stopping). The model parameters and data augmentations were initialized with random seed 2060. To account for stochastic training process, we trained each model two more times with random seeds 12060 and 22060. 

We used  binary cross entropy loss for training the neural network, defined as
\begin{equation}
    l = \frac{1}{N\times w \times h} \sum_{i=0}^{N\times w \times h} y_{i} \log (\sigma (x_{i})) + (1-y_{i}) \log(1-\sigma (x_{i})),
\end{equation}
where $N$ is the batch size, $w, h$ are width and height of the 2D projection image, $x_{i}$ is the network output at pixel location $i$, $\sigma$ is the sigmoid function, and $y_{i}$ is the corresponding ground truth pixel value.

\subsection{Evaluation}
We used 100 manually segmented metals from projections of 10 clinical scans as ground truths for evaluation of different metal segmentation methods. We evaluated U-Net-based models trained on noisy and clean simulations. For each of the noisy and clean simulations, the model was trained on full size, crops, and their combination. For forward projection-based conventional MAR (CMAR), we segmented the metals in 3D images using connected component region growing segmentation similar to \cite{Mey:10}. Metal seeds were started at the voxels with intensities of more than 7000 HU. The metal region was grown starting from the seeds, by checking the intensities of the connected pixels in the neighborhood. If the neighboring pixel had an intensity of more than 3000 HU, it was included in the metal region. The segmented 3D metals were forward projected to create metal traces in the 2D projections. We also experimented with higher values than 3000 HU for lower (region growing) threshold, but then some metals were missed and undersegmented.

We also compared our results with moving metal artifact reduction (MMAR) \cite{hahn18}. We used MMAR to refine the coarse metal traces obtained by CMAR.
For the patch-based training (Noisy (crops) and Clean (crops)), we followed \cite{Got21} and used random crops of size $512 \times 512$ during the training. The inference was done on full size projections. For a fair comparison, we used the same U-Net network with the same parameters for all compared methods. For the comparison with the consistency check, we followed \cite{Got21} to reconstruct 3D binary metal from the initial segmentation of Noisy (full+crops) model and forward project 3D binary metal to get updated metal traces. 

The quantitative segmentation performance of the compared methods was evaluated using Dice Similarity Coefficient (DSC), Intersection over union (IOU) and False positive rate (FPR) \cite{taha_15}. These metrics are defined as

\begin{align} 
    \textrm{DSC} &=  \frac{2 \textrm{TP}}{2 \textrm{TP} + \textrm{FP} + \textrm{{FN}}}, \\
    \textrm{IOU} &=  \frac{\textrm{TP}}{\textrm{TP} + \textrm{FP} + \textrm{{FN}}}, \\
    \textrm{FPR} &= \frac{\textrm{FP}}{\textrm{TN}},
\end{align} 
where TP is true positive, TN is true negative, FP is false positive, and FN is false negative.

 Metal traces found from the compared methods were inpainted from the nearby pixels using a 2D interpolation method. The inpainted projections were reconstructed using a modification of FDK \cite{Feldkamp:84} algorithm. The 3D metals obtained from the region growing-based method were put back in the reconstructions.
 \begin{table}[hbt!]
\caption{Mean DSC (\%) and mean IOU (\%) ( $\pm$ standard error) calculated for metal trace segmentation on 10 test scans. $\uparrow$ indicates that higher values are better.}

\centering
\small
    \begin{tabular}{cccccc}
\hline
{Method} & {mean DSC {$\uparrow$}} & {mean IOU $\uparrow$} \\
\hline
{CMAR} & $90.5 \pm 2.0 $ & $84.4 \pm 3.2 $\\
\hline
{MMAR} & $89.1 \pm 2.2$ & $81.5 \pm 3.4$ \\ 
\hline
{Noisy (full)} & $93.7 \pm 0.9$ & $88.6 \pm 1.6$  \\
\hline
{Noisy (crops)} & $94.4 \pm 0.7$ & $89.6 \pm 1.2$ \\
 \hline
{Noisy (full+crops)} & $\mathbf{94.8} \pm \mathbf{0.6}$ & $\mathbf{90.2} \pm \mathbf{1.1}$ \\
\hline
{Consistency check} & $91.0 \pm 3.4$ & $85.0 \pm 4.8$ \\
 \hline
 {Clean (full)} & $93.0 \pm 1.0 $ & $87.4 \pm 1.6$ \\
  \hline
 {Clean (crops)} & $92.2 \pm 1.6 $ & $86.5 \pm 2.1$ \\
 \hline
 {Clean (full+crops)} & $93.8 \pm 0.8$ & $88.6 \pm 1.4$ \\
 \hline
\end{tabular}
\label{table4}
\end{table}
\section{Results}
In this section we compare the quantitative metal segmentation performance of U-Net-based models with the forward projection-based CMAR, MMAR, and the consistency check. Then we show the qualitative impact of metal trace segmentation on the reconstructed images.

\subsection{Quantitative Analysis}
\begin{figure}
\centering
\includegraphics[width=\columnwidth]{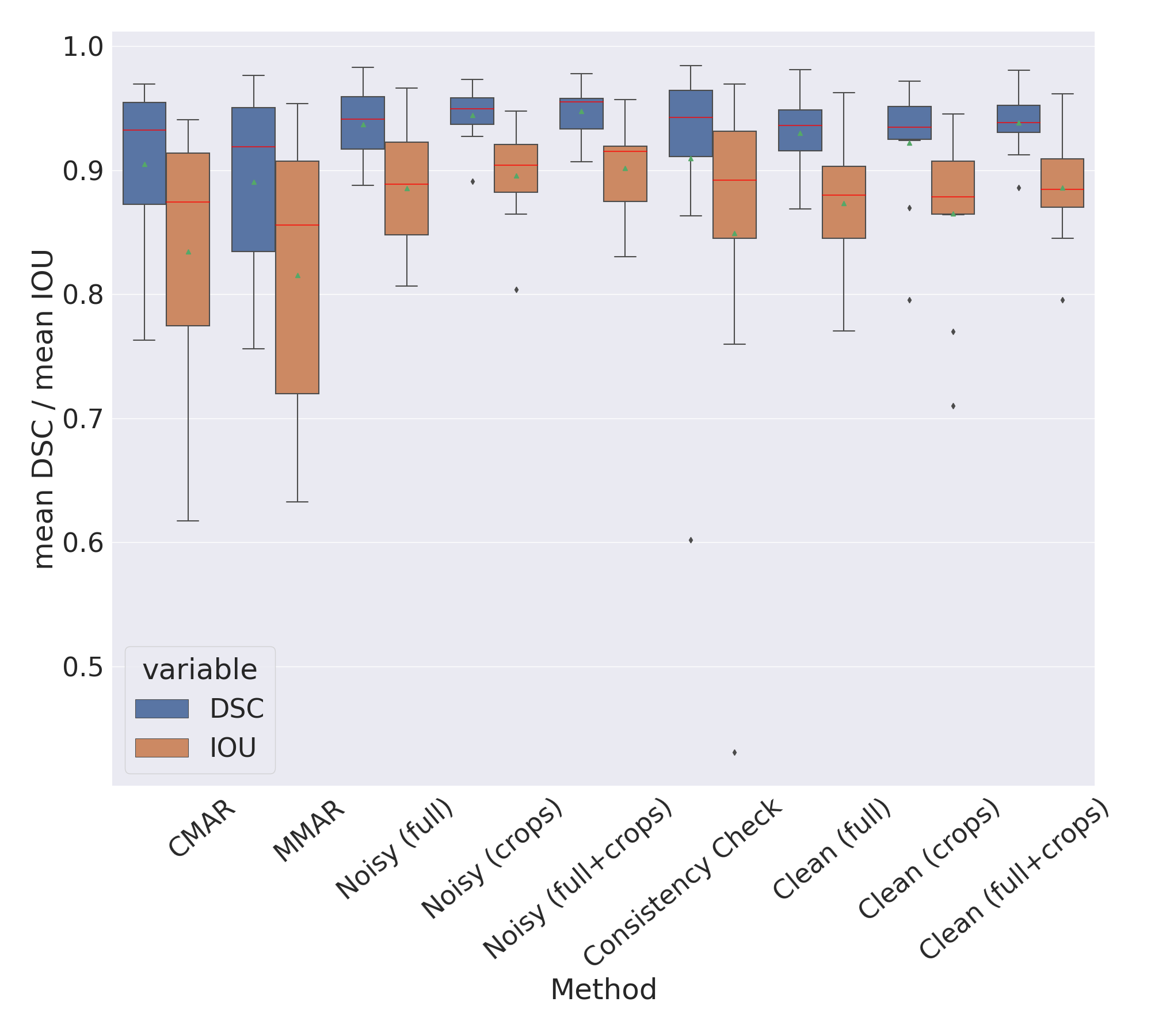}
\caption{Boxplot visualization of DCE and IOU scores for CMAR, MMAR, U-Net models, and consistency check. The rectangles contain data within the first and the third quartiles. The endpoints of the lower and upper whiskers are selected as the first quartile - 1.5 times the interquartile range (IQR) and third quartile + 1.5 IQR,  respectively. The medians are visualized as red lines and the means as green triangles. The outliers are the points that are outside the interval defined by the whiskers.}
\label{boxplot}
\end{figure}

\subsubsection{Clinical Data with Metals}
As performance measures, we calculated the mean DSC and IOU for 10 test scans. Table \ref{table4} shows the segmentation performance of CMAR, MMAR, six U-Net-based models, and the consistency check-based segmentation applied on the model Noisy (full+crops). Compared to the CMAR and MMAR, all six U-Net-based models had higher mean DSCs and IOUs. This shows that the deep learning methods segmented metals more accurately than the forward projection-based CMAR and MMAR. The model Noisy (full+crops) gave the highest mean DSC=94.8 (SE=0.6) and IOU= 90.2 (SE=1.1). When MMAR was applied on the dilated metal traces obtained from CMAR, the mean DSC decreased from 90.5 (SE=2.0) to 89.1 (SE=2.2) and the mean IOU decreased from 84.4 (SE=3.2) to 81.5 (3.4). Furthermore, when consistency check \cite{Got21} was applied on the best model Noisy (full+crops), it decreased the mean DSC from 94.8 (SE=0.6) to 91.0 (SE=3.4) and the mean IOU from 90.2 (SE=1.1) to 85.0 (SE=4.8). The mean DSC and IOU scores decreased due to the challenging cases of metals which is explained further in the qualitative analysis in Section \ref{ssec:qualitative}. Among the three models trained on clean simulations, Clean (full+crops) had the highest mean DSC=93.8 (SE=0.8) and IOU=88.6 (SD=1.4), which is similar to the finding that the Noisy (full+crops) model had the highest mean DSC and IOU among the three models trained on noisy simulations.

We present the DSC and IOU results in Fig.~\ref{boxplot}. The box plots of DSCs and IOUs of MMAR had the largest spread over values. The median values of DSCs and IOUs of Noisy (full + crops) model were close to the median values of Noisy (crops). Boxplots of Noisy (full+crops) did not have any outliers as opposed to one outlier in Noisy (crops). The boxplots of Clean (crops) and Clean (full+crops) had more outliers than Noisy (crops) and Noisy (full+crops). Furthermore, the boxplots of consistency check had more vertical spread and showed one outlier in comparison to the boxplots of Noisy (full + crops).

\begin{figure*}
\centering
\includegraphics[width=.9\textwidth]{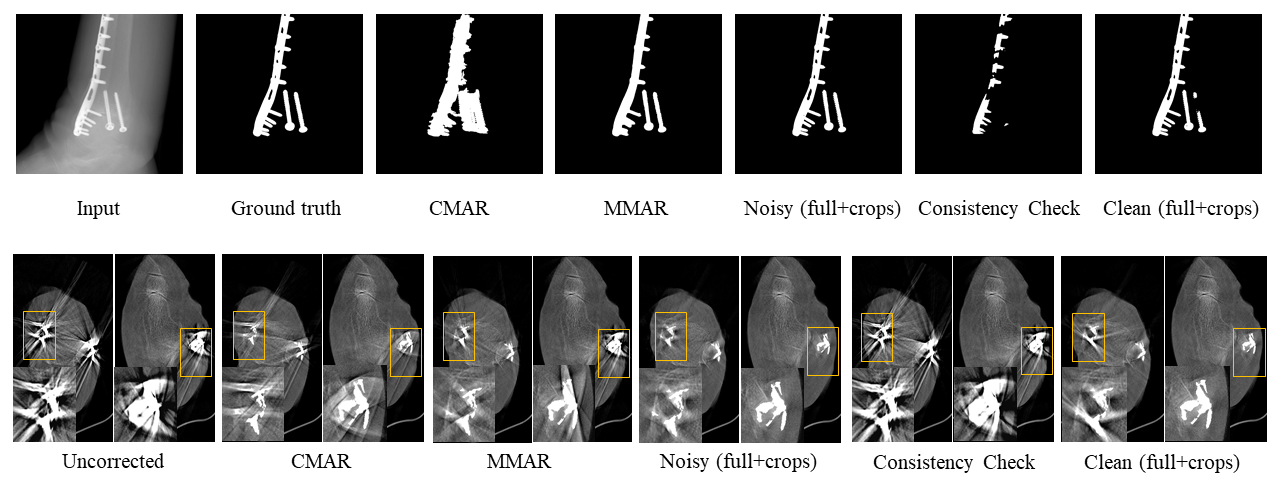}
\caption{Test scan 4 with large motion and complex metal implants. The first row shows segmentation of metals in an input projection by various methods. The second row shows two axial image slices from the reconstructions (slice location 370 and 511, respectively) and the corresponding magnified region of interest. Yellow boxes show the location of the region of interest in the axial image. Our method, Noisy (full+crops) segmented metals in the projection most accurately which reduced most of the artifacts in the reconstructed axial slices.}
\label{scan4}
\end{figure*}
\subsubsection{Clinical Data without metals}
When comparing the segmentation performance of the models in the presence of the metals, it is also important to know how the models behave when the CBCT projections do not contain any metal. This comparison might show the model's generalization. This comparison does not require ground truth metal traces. Table \ref{table5} shows the FPR of the 6 models calculated for metal-free projections from 6 clinical scans. The mean FPR for the Noisy (full+crops) was 0.51$\times 10^{-3}$ (SE=0.16$\times 10^{-3}$) which was less than the mean FPR given by Noisy (crops), i.e., 0.85$\times 10^{-3}$ (SE=0.36$\times 10^{-3}$) and the mean FPR given by Noisy (full), i.e., 14$\times 10^{-3}$ (SE=5$\times 10^{-3}$). Similarly, the model trained on simulations with 5000 photons per pixel (Clean (full+crops)) had least mean FPR i.e., 0.44$\times 10^{-3}$ (SE=0.16$\times 10^{-3}$) compared to the mean FPR given by Clean (crops), i.e., 8$\times 10^{-3}$ (SE=2$\times 10^{-3}$) and the mean FPR given by Clean (full), i.e., 23$\times 10^{-3}$ (SE=5$\times 10^{-3}$). Overall, this analysis on the metal-free projections showed that the models trained on the full size and cropped projections together were more robust the false positives compared to the models trained on only full size or only cropped projections.
\begin{table}[hbt!]
\caption{Mean FPR ($10^{-3}$) ($\pm$ standard error) calculated for 6 scans without any metals. $\downarrow$ indicates that lower values are better.}
\centering
\small
    \begin{tabular}{cccc}
\hline
{Method} & FPR $\downarrow$ \\ 
\hline
{Noisy (full)} & $ 14 \pm 5  $ \\
 \hline
 {Noisy (crops)} & $0.85 \pm 0.36 $\\
 \hline
{Noisy (full+crops)}  & $0.51 \pm 0.16 $ \\
 \hline
{Clean (full)}  & $23 \pm 5 $\\
 \hline
{Clean (crops)}  & $8 \pm 2 $ \\
 \hline
{Clean (full+crops)}  & $\mathbf{0.44 \pm 0.16} $ \\
 \hline
\end{tabular}
\label{table5}
\end{table}

\subsubsection{Comparison of different training runs} The comparison of DSC scores on test data for the 6 models obtained from 3 independent training runs is shown in Table~\ref{dsc}. The mean DSC score for each run is shown in the columns named, seed1 (2060), seed2 (12060), and seed3 (22060). The last column named delta is the difference between maximum and minimum DSC scores across three runs. The model Noisy (full+crops), trained on the noisy full size and cropped projections, had the best mean DSC scores in each independent run, i.e., 94.8 (SE=0.6), 94.5 (SE=0.7), and 94.5 (SE=0.7). The delta for the model Noisy (full+crops) was 0.3, which was less than the delta of model Noisy (full), i.e., 2.3 and the delta of model Noisy (crops), i.e., 1.9. Similarly, the delta for the model Clean (full+crops) trained on clean full size and crops was 0.4, which was less than the delta of model trained on Clean (full), i.e., 2.1 and the delta of the model Clean (crops), i.e., 2.4. It can be deduced that the models trained on full size and crops were giving more stable mean DSC scores across three independent runs. Furthermore, the model Noisy (full+crops) had a slightly smaller delta (0.3) than the delta of the model Clean (full+crops), i.e., 0.4. These results show that the model trained on noisy full size and cropped projections was most robust model in terms of DSC scores.

\begin{table}[hbt!]
\caption{Mean DSC (\%) ($\pm{}$ standard error) calculated for three training runs for metal trace segmentation on 10 test scans. Delta is difference between maximum and minimum mean DSC for the corresponding method.}

\centering
\setlength{\tabcolsep}{0pt} 
\begin{tabular*}{\columnwidth}{@{\extracolsep{\fill}\quad}cccccc}
\hline
{Method} & {seed 1} & {seed 2} & {seed 3} & delta \\ 
\hline
{Noisy (full)} & $93.7 \pm 0.9 $ & $92 \pm 1.4 $ & $94.2 \pm 0.8 $  & $2.2$ \\
\hline
{Noisy (crops)} & $94.4 \pm 0.7 $ & $92.5 \pm 1.6 $ & $92.5 \pm 1.3 $ & $1.9$ \\
 \hline
{Noisy (full+crops)} & $\mathbf{94.8} \pm \mathbf{0.6}$ & $\mathbf{94.5} \pm \mathbf{0.7}$ & $\mathbf{94.5} \pm \mathbf{0.7}$ & $\mathbf{0.3}$ \\
\hline
 {Clean (full)} & $93.0 \pm 1.0 $ & $90.9 \pm 2.2$ & $91.9 \pm 1.4$ & $2.1$ \\
 \hline
 {Clean (crops)} & $92.2 \pm 1.6 $ & $91.2 \pm 2.0 $ & $93.4 \pm 1.0 $ & $2.4$ \\
 \hline
 {Clean (full+crops)} & $93.8 \pm 0.8 $ & $93.4 \pm 0.9$ & $93.4 \pm 1.1$ & $0.4$ \\
 \hline
\end{tabular*}
\label{dsc}
\end{table}

\begin{figure*}[!t]
\centerline{\includegraphics[width=\textwidth]{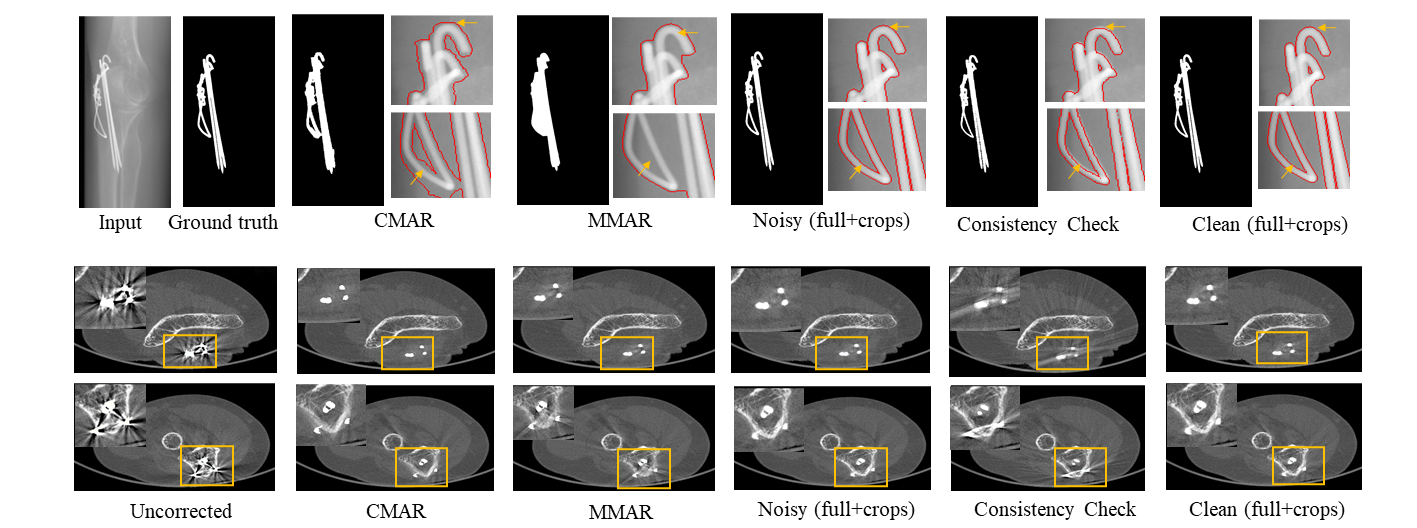}}
\caption{Test scan 9 with a complex metal shape. The first row shows an input projection, ground truth, and segmentation of the metal by several methods. Parts of the metal are enlarged to show the minute segmentation differences on the right of the segmented projections. The second and third rows show two axial image slices from the reconstructions with the corresponding magnified regions of interest. The axial slices are from slice location 169 and 396, respectively. Yellow boxes show the location of the region of interest in the axial image. Noisy (full+crops) and Clean (full+crops) segmented metals well which reduced most of the artifacts and preserved more details in the reconstructions.}
\label{scan9}
\end{figure*}

\subsection{Qualitative Analysis}
\label{ssec:qualitative}
For the qualitative analysis, we show the metal segmentation and the effect of accurate metal segmentation on the corresponding reconstructions. For all qualitative visualizations we compared Noisy (full+crops) reconstructions with the reconstructions from CMAR, MMAR, consistency check step \cite{ketcha21} applied on the segmentation from Noisy (full+crops), and Clean (full+crops). We analyzed improvements in the image quality in complex cases such as the motion affected and out-of-FOV metal cases.

\begin{figure*}[!htb]
\centerline{\includegraphics[width=\textwidth]{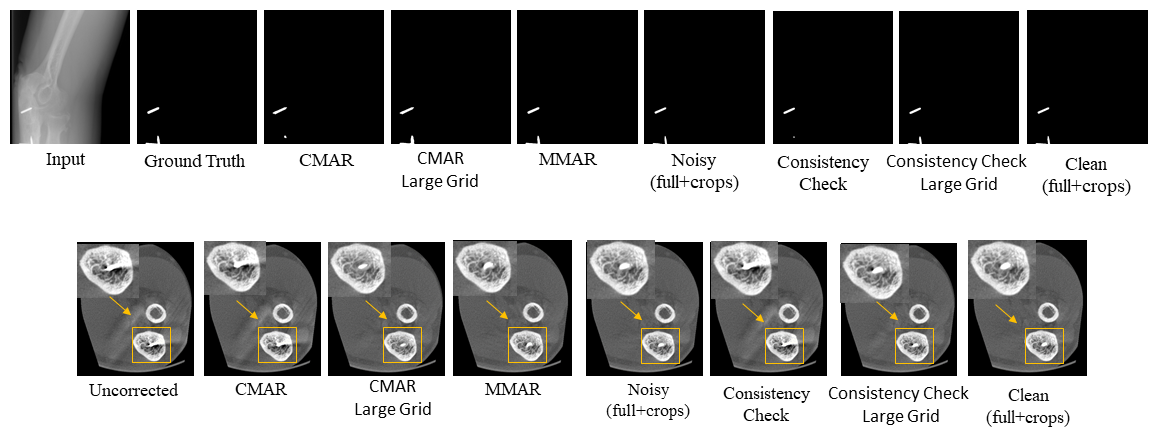}}
\caption{Test scan 6 containing out-of-FOV metal. The first row shows segmentation of the metals in the input projection by different methods and corresponding grond truth. The second row shows an axial slice from the reconstruction with the corresponding magnified region of interest. MMAR, Noisy (full+crops), and Clean (full+crops) segmented metals accurately without the need of larger reconstruction.}
\label{scan6}
\end{figure*}

\begin{figure*}[!t]
\centerline{\includegraphics[width=\textwidth,height=9cm,keepaspectratio]{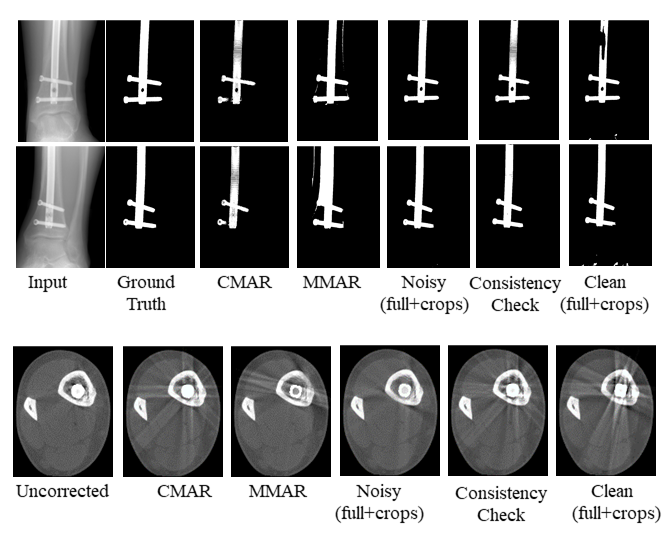}}
\caption{Test scan 8 with 0.6 mm voxel size. The first and second rows show segmentation of metals in the input projections by different methods and the corresponding ground truths. The third row shows an axial slice from the reconstructions. Our method, Noisy (full+crops) segmented metal traces best.}
\label{scan8}
\end{figure*}

\subsubsection{Motion affected volume}
The qualitative results on the challenging motion case are shown in the Fig.~\ref{scan4}. The segmentation of metals and two axial images from the reconstructions are visualized. The uncorrected images contained large artifacts from the presence of the large metals as well as the motion during the scan. The CMAR method over-segmented the metals in the projection. The conventional method reduced some artifacts in the reconstructed images but still there were many remaining streaks and darkening around the metal. The over-segmentation of metals in CMAR reduced some bone details in the axial image as well. MMAR refined the segmentations from CMAR but the holes inside the metals could not be recovered. The corresponding image slices for MMAR still had some remaining artifacts. The Noisy (full+crops) model segmented metals in projections most accurately. Most of the artifacts were reduced and the bones near the metal were clearly visible in the axial images. Because of the large motion, the consistency check method missed some of the blurred metals in the reconstructions and the forward projection of these metals removed most of the metal trace in the projections. This caused the artifacts to reappear in the images and affected image quality adversely. The model Clean (full+crops) missed some part of metals in the projection, this caused artifacts in one of the axial images. 

\subsubsection{Under- and over-segmentation}
Another interesting case is shown in Fig.~\ref{scan9}. The metals were over-segmented by the region growing-based CMAR. The over-segmentation reduced some details near the metal such as the bone is blurred in the reconstructed axial image. MMAR could not deal with the complex shape of the metal and dilated the CMAR segmentation more. On the other hand, Noisy (full+crops) and Clean (full+crops) models segmented the metal traces well. As a result, the axial images from both of those methods had more details visible around the metal and the bone is better preserved. The consistency check missed only small part of the metal traces in the projection. The small under-segmentation caused significant artifacts in the reconstructed images. The results clearly showed that the under-segmentation may cause large artifacts in complex cases and over-segmentation may blur crucial details. 

\subsubsection{Out-of-FOV metal}
Fig.~\ref{scan6} shows when metal was in the out-of-FOV region, forward projections of image-domain metal failed to segment metal traces in the projection domain (CMAR and consistency check), which, in turn, caused artifacts in the reconstructions. MMAR was able to recover the metals from the CMAR and did not need larger reconstruction. The original volume size for the reconstruction of the scan was $801 \times 801 \times 601$. To account for out-of-FOV metals in the forward projections, image reconstruction needed to be done on a larger size grid. When reconstruction was done for a volume size of $801 \times 801 \times 661$, the forward projection included all of the metals in the projections (CMAR large grid and consistency check large grid). Noisy (full+crops) and Clean (full+crops) models segmented the metal traces directly in the projections without the need of larger reconstruction. As we used the same metal inpainting and reconstruction procedure for all methods, it is clear from Fig.~\ref{scan6} that most of the artifacts were reduced in the reconstructed images if the metals were segmented well in the projections. In this case, the out-of-FOV metal was relatively close to the reconstruction edge. In some clinical cases where the out-of-FOV metal is far from the reconstruction FOV, the segmentation of metal is not possible. In such situations, our method will still segment metals in the projections without large reconstructions.

\subsubsection{Effect of voxel size}
The effect of small voxel sizes on the forward projections of metals is shown in Fig.~\ref{scan8}. The images were reconstructed with a voxel size of 0.6 mm. The forward projection did not consider the voxel size and sampling artifacts are visible in the metal projections of the CMAR. Furthermore, CMAR missed some part of the metal, which was recovered by consistency check and MMAR. However, the MMAR segmentation resulted in oversegmentation of the metals and the consistency check segmentation still shows some sampling artifacts. Additional artifact can be seen in the reconstruction of CMAR, MMAR, and the consistency check. The model trained on the clean simulations missed some metal part in the projections which caused artifacts in the reconstructed images. The model trained on noisy simulations segmented metals well which did not cause any additional artifacts. This visualization shows, that our method does not need to account for the voxel size as it segments metals directly in the projections.

\section{Conclusion}
In this paper, we proposed to use noisy Monte Carlo simulations to train a U-Net architecture for the segmentation of metals in CBCT projections. We experimentally demonstrated that synthetic data could substitute real data for metal segmentation training. We showed that the model trained with noisy simulations outperformed the model trained on time-consuming cleaner simulations for the metal segmentation task. We also showed that adding the crops to the full size projections during the training helped to get more robust metal segmentation on the real clinical test scans. The forward projection based methods, such as CMAR, MMAR, and consistency checks are prone to errors and require tuning of parameters. However, segmenting the metals directly in projections is more robust solution.

The model trained on full size projections and crops, significantly reduced the false positives in the non-metal projections. Although, we trained our model on limited simulations, we have shown the applicability of the model on a diversified real dataset acquired from multiple anatomies and scanner sites. We have discussed noticeable improvements in the reconstruction image quality of motion-affected and out-of-FOV metal affected scans by obtaining better segmentation traces of metals. Recently, a Fourier dual-domain restoration network has been shown to be more robust to metal trace defects \cite{li22} which is another approach to alleviate the artifacts coming from wrong segmentation of metals. In our future work, we will consider to test the application of our method on more diversified and larger dataset including dental scans.

\section{Study Limitations}
While we have demonstrated excellent results for metal artifact correction using our method, we have used only simulated data for the training. This approach helps to address the challenges associated with obtaining enough clinical data for training and manually segmented ground truth. However, simulated projections might not fully capture all the variations and complexities inherent in the clinical scans. This could potentially impact the  generalizability and effectiveness of our proposed method in real-world scenarios. To address this limitation, we incorporated noisy simulations and included 10 clinical scans for testing. Although we have included complex cases, such as those with motion artifact and out-of-fov metals, there is still a necessity to include a larger dataset to fully establish the model's generalizability and effectiveness.

As we have noted earlier, metal artifact reduction methods often require precise segmentation of metal traces in CBCT projections. The efficacy of the trained model's performance depends upon the accuracy of this ground truth metal segmentation. If the metals in the simulated training data are not segmented accurately, the performance of the trained model may degrade on the real clinical cases.

\bibliographystyle{IEEEtran}
\bibliography{refs}
\begin{IEEEbiography}
[{\includegraphics[width=1in,height=1.25in,clip,keepaspectratio]{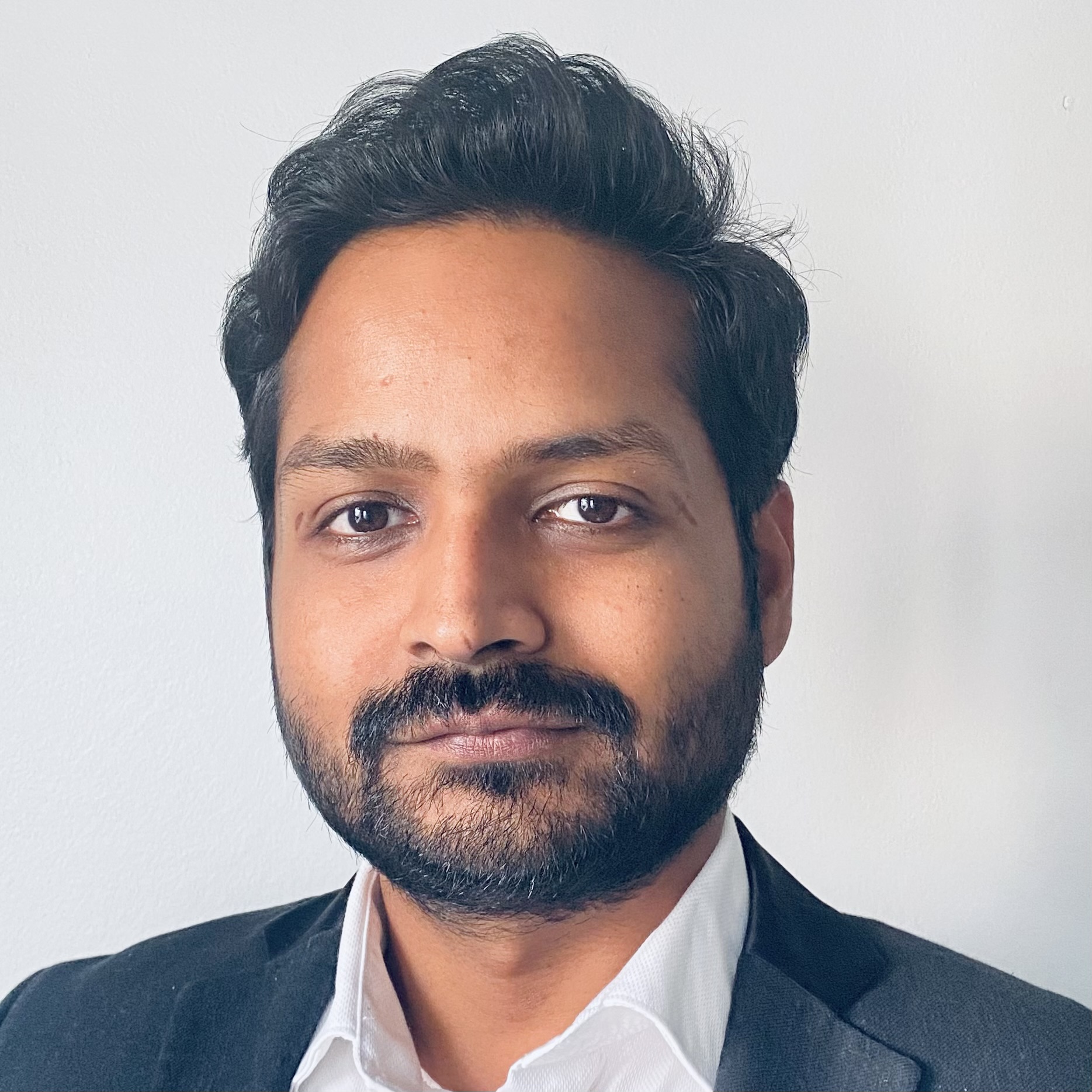}}]{Harshit Agrawal} received the B.Tech. degree (Gold medalist) from U.P.T.U., India, in 2011, and the M.Sc. (Tech.) degree from Aalto University, Finland, in 2018, where he is currently pursuing the D.Sc. degree with the Department of Electrical Engineering and Automation Engineering. He has worked as Senior Engineer with GE healthcare, India, and is currently working as Machine Learning Specialist with Planmeca Oy., Finland. His research interests include machine learning, especially application of deep neural networks in medical imaging.
\end{IEEEbiography}

\begin{IEEEbiography}
[{\includegraphics[width=1in,height=1.25in,clip,keepaspectratio]{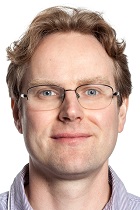}}]
{Ari Hietanen} received his Ph.D. degree in theoretical physics from the university of Helsinki, Finland, in 2008. He is currently research manager in algorithms at Planmeca Oy. His current research interests include machine learning and medical physics. 
\end{IEEEbiography}
\begin{IEEEbiography}
[{\includegraphics[width=1in,height=1.25in,clip,keepaspectratio]{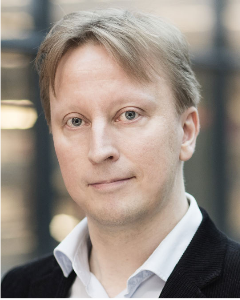}}]
{Simo S\"{a}rkk\"{a}} (Senior Member, IEEE) is currently an Associate Professor with Aalto University, Finland. He has authored or coauthored over 100 peer-reviewed scientific articles and three books. His research interests include multi-sensor data processing systems and machine learning methods with applications in medical and health technology, target tracking, inverse problems, and location sensing. He is a member of the IEEE Machine Learning for Signal Processing Technical Committee. He has been serving as an Associate Editor for the IEEE Signal Processing Letters.
\end{IEEEbiography}
\EOD

\end{document}